\documentclass[aps,pra,showpacs,floatfix,nofootinbib,superscriptaddress]{revtex4}
\usepackage{graphicx} 
\usepackage{amsmath}
\usepackage{bm}
\usepackage{dcolumn}

\begin{document}

\title{Dynamic polarizabilities and
related properties of clock states of ytterbium atom}

\author{V. A. Dzuba}
\affiliation {School of Physics, University of New South Wales, Sydney,
2052, Australia}

\author{A. Derevianko }
\affiliation{Physics Department, University of
Nevada, Reno, Nevada  89557, USA}


\date{\today}
\begin{abstract}
We carry out relativistic many-body calculations of the static and
dynamic dipole polarizabilities of the ground $6s^2\, ^1\!S_0$ and the first excited $6s6p\, ^3\!P^o_0$ states of
Yb. With these polarizabilities, we compute several properties of Yb
relevant to optical lattice clocks operating on the $6s^2\, ^1\!S_0 - 6s6p\, ^3\!P^o_0$  transition.
  We determine (i) the first four {\em magic} wavelengths of
the laser field for which the frequency of the clock transition is
insensitive to the laser intensity. While the first magic wavelength is known, we predict
the second, the third and the forth magic wavelengths to be 551 nm, 465 nm, and 413 nm.
(ii) We reevaluate  the effect of black-body radiation on the frequency of the clock transition,
the resulting clock shift at $T=300 \, \mathrm{K}$ being $-1.41(17)$~Hz.
(iii) We compute long-range interatomic van der Waals coefficients (in a.u.)
$C_6(6s^2\, ^1\!S_0 +6s^2\, ^1\!S_0) = 1909(160)$, $C_6(6s^2\, ^1\!S_0 + 6s6p\, ^3\!P_0) =2709(338) $, and
$C_6(6s6p\, ^3\!P_0 + 6s6p\, ^3\!P_0) =3886(360) $. Finally, we  determine the atom-wall interaction
coefficients (in a.u.), $C_3 (6s^2\, ^1\!S_0) =3.34$ and $C_3 (6s6p\, ^3\!P_0) =3.68$.
We also address and resolve a disagreement between previous
calculations of the static polarizability of the ground state.
\end{abstract}

\maketitle

\section{Introduction}

Ytterbium atom is employed in a number of projects aimed at studying fundamental
problems of modern physics by means of atomic physics. A large parity-violating
signal on the $6s^2 \, ^1\!S_0$ - $6s5d \ ^3\!D_1$ 408-nm
transition of Yb has been recently observed at Berkeley~\cite{TsiDouFam09}.
A search for the CP-violating (T-, P- odd)
permanent electric dipole moment of Yb was initiated~\cite{YbEDM}. A Bose-Einstein condensate
in a dilute gas of the ground-state Yb atoms has been attained~\cite{TakMakKom03} and a
number of experiments with ultracold Yb atoms has been carried out (see, e.g., Refs.~\cite{EnoKasKit08,KitEnoKas08,FukSugTak09}).

Here we focus on the optical lattice clock applications of  Yb.
In the optical lattice clocks, neutral atoms are trapped in a
standing wave of a laser light (optical lattice) operated at a
certain ``magic'' wavelength~\cite{KatTakPal03}. At this
wavelength a differential light
perturbation of the two clock levels vanishes exactly. Using Yb atom
in lattice clocks was proposed in Ref.~\cite{PorDerFor04} and
experimentally demonstrated in Boulder~\cite{BarHoyOat06,LemLudBar09}.
Apart from numerous technical applications, such atomic
clocks may be used for searching for a variation of fundamental constants~\cite{FlaDzu09},
mapping out atom-wall interaction~\cite{DerObrDzu09}, and
serve as a basis for quantum information processing (see, e.g., \cite{HayJulDeu07,DalBoyYe08}).

In the present paper, we evaluate dynamic electric-dipole polarizabilities of the clock states, $6s^2 \, ^1\!S_0$ and $6s6p \, ^3\!P^o_0$.
Dynamic polarizabilities of real and imaginary frequencies are linked to
a variety of important atomic and molecular properties.
In particular, we compute magic wavelengths,
the effect of black-body radiation on the frequency of the
clock transition, atom-wall interaction constants $C_3$, and, finally, long-range molecular van der
Waals coefficients $C_6$ for the clock states.  We take advantage of a close link
between these properties and improve our theoretical predictions by
referencing to measured properties, such as the first magic wavelength~\cite{BarHoyOat06,BarStaLem08} and the van
der Waals coefficient for two ground-state atoms~\cite{KitEnoKas08}.

Ytterbium has been studied both experimentally (see, e.g.,
 \cite{TakKomHon04,FukSugTak09,BarHoyOat06,EnoKitKas07,BarStaLem08,PolBarLem08,KitEnoKas08,LemLudBar09})
 and theoretically (see, e.g.,
\cite{PorDerFor04,PorRakKoz99P,PorDer06Z,PorDer06BBR,OvsPalTai07,ChuDalGro07,ZhaDal07,
ZhaDal08,SahDas08}.
Trapping and cooling schemes were discussed and implemented, the frequency of
the clock transition was measured to a high accuracy~\cite{LemLudBar09}, the first magic
wavelength~\cite{BarHoyOat06,BarStaLem08}
and the van der Waals coefficient for the ground state of Yb were also
measured~\cite{KitEnoKas08}.

Theoretical studies are not free from a controversy.
Calculated polarizability of the Yb ground state  reported in
Ref.~\cite{PorDer06Z} is in a
disagreement with most of other {\em ab initio} or semi-empirical analyses. The
source of this disagreement is explained below and the discrepancy is
resolved. The resolution is closely related to the role of the
excitations from the core which are discussed in detail.
Corresponding statement
can be formulated as a general theorem:
{\em   Consider a second-order transition matrix element, involving summation
over a complete set of intermediate states. Then,
a contribution from a subspace spanned by degenerate
states does not depend on mixing of these states}.  This statement has implications
not only for computing polarizabilities,  but also for the
two-photon transition amplitudes, parity-violating amplitudes, etc. in Yb and other
atoms.

\section{Method of calculation}
\label{method}
In this section we start by recapitulating expressions for dynamic polarizability and
its relations to magic wavelengths, and atom-atom and atom-wall interaction coefficients.
Next we discuss our computational relativistic many-body method.

The ac Stark shift of an energy of a spherically-symmetric, $J=0$, atomic state $v$ in electro-magnetic wave of amplitude
$\varepsilon$ and frequency $\omega$ is given by
\begin{equation}
  \Delta E_v(\omega) = -\alpha_v(\omega)\left(\frac{\varepsilon}{2}\right)^2,
\label{DeltaE}
\end{equation}
where  the electric-dipole dynamic polarizability
reads (we use atomic units in which $\hbar=1$, $|e|=1$, and $m_e=1$)
\begin{equation}
  \alpha_v(\omega) = \frac{2}{3}\sum_k
  \frac{E_k-E_v}{(E_k-E_v)^2-\omega^2}\langle v||\mathbf{d}||k\rangle^2 \, .
\label{alphaw}
\end{equation}
Here $\mathbf{d}$ is an electric dipole operator and summation is
over a complete set of many-body states.

{\em Magic} laser frequency $\omega^*$  is
found from the condition that the ac Stark shifts of both clock states $v$ and
$w$ are identical so that the
frequency of the transition does not depend on the laser intensity,
\begin{equation}
  \Delta E_v(\omega) - \Delta E_w(\omega) =
  -(\alpha_v(\omega)-\alpha_w(\omega)) \left(\frac{\varepsilon}{2}\right)^2=0.
\label{omega*}
\end{equation}
Simply, at the magic frequencies, $\alpha_v(\omega^*)=\alpha_w(\omega^*)$.

Clock frequency shift due to  black-body radiation is expressed in terms of the difference of static polarizabilities of the two clock levels (see detailed discussion in Ref.~\cite{PorDer06BBR}).

We also require atomic polarizabilities evaluated at purely imaginary frequencies
\begin{equation}
  \alpha_v(i\omega) = \frac{2}{3}\sum_k
  \frac{E_k-E_v}{(E_k-E_v)^2+\omega^2}\langle v||\mathbf{d}||k\rangle^2.
\label{alphaiw}
\end{equation}
These are used in evaluating the atom-wall interaction constant
$C_3$~\cite{DerJohSaf99}
\begin{equation}
  C_3 = \frac{1}{4\pi}\int_0^{\infty} \alpha(i\omega) d\omega,
\label{eq:C3}
\end{equation}
and the van der Waals coefficients $C_6$.  For two separated atoms in states $w$ and $v$,
\begin{equation}
  C_6 (w+v) = \frac{3}{\pi}\int_0^{\infty} \alpha_w(i\omega) \alpha_v(i\omega) d\omega \, .
\label{eq:C6}
\end{equation}
The $C_6$ for two atoms interacting in identical states is subsumed by the above formula. Eq.(\ref{eq:C6}) is derived under assumption  that there are no downward dipole transitions from either $w$ or $v$ state, which holds in our case.

It is clear that for computing the polarizabilities,  we need to determine electric dipoles and energies, and to
perform the summation over a complete set of atomic states. To this end,
we use a combination of the configuration-interaction (CI) method and the
many-body perturbation theory (MBPT) to construct an effective
Hamiltonian for two valence electrons (the CI+MBPT
method~\cite{DzuFlaKoz96b,DzuJoh98}). Further, we employ the Dalgarno-Lewis
method~\cite{DalLew55} to reduce the summation over a complete set of
many-electron states to solving an inhomogeneous system of linear equations.

\subsection{CI+MBPT method}

The ground state configuration of Yb is $4f^{14}6s^2$ and most of the
low-energy excited states correspond to configurations with one of the $6s$ electrons
being excited. All these states can be represented to a relatively high accuracy
as states with two valence electrons above a closed-shell
core. This is the starting point of our calculations: the computational (CI) model space is spanned by all possible excitations of the two valence electrons. Implications arising from the states with excitations from the outer
$4f$ core sub-shell will be discussed below.

The effective CI+MBPT Hamiltonian for two valence electrons has the form
\begin{equation}
  \hat H^{\rm eff} = \hat h_1(r_1) + \hat h_1(r_2) + \hat h_2(r_1,r_2),
\label{Heff}
\end{equation}
where $\hat h_1$ is the single-electron part of the relativistic Hamiltonian
\begin{equation}
  \hat h_1 = c \mathbf{\hat{\alpha}} \mathbf{p} + (\hat{\beta}-1)m_e c^2-\frac{Ze^2}{r}
  + V^{N-2} + \hat \Sigma_1,
\label{h1}
\end{equation}
and $\hat h_2$ is the two-electron part of the Hamiltonian
\begin{equation}
  \hat h_2(r_1,r_2) = \frac{e^2}{|\mathbf{r}_1 - \mathbf{r}_2|} + \hat
  \Sigma_2(r_1,r_2).
\label{h2}
\end{equation}
In these equations, $\mathbf{\hat{\alpha}}$ and $\hat{\beta}$ are the conventional Dirac matrices,
$V^{N-2}$ is the Dirac-Hartree-Fock (DHF) potential of the closed-shell atomic
core ($N-2=68,Z=70$), and $\hat \Sigma$ is the correlation operator. It
represents terms in the Hamiltonian arising due to virtual excitations from
atomic core (see Ref.~\cite{DzuFlaKoz96b,DzuJoh98} for details).
$\hat \Sigma \equiv 0$ corresponds to the standard CI method.
$\hat \Sigma_1$ is a single-electron operator. It represents a
correlation interaction (core-polarization) of a particular valence electron with the atomic
core. $\hat \Sigma_2$ is a two-electron operator. It represents
screening of the Coulomb interaction between the two valence electrons by the core
electrons. We calculate $\hat \Sigma$ in the second order of the
MBPT. We use a  B-spline technique~\cite{JohSap86} to construct a
complete set of single-electron orbitals. We use 40 B-splines in a
cavity of radius $R=40\, a_B$ and calculate the eigenstates of the $V^{N-2}$
DHF Hamiltonian up to the maximum value of the angular
momentum $l_{max}=5$. The same basis is used in computing $\hat
\Sigma$ and in constructing the two-electron states for the valence
electrons. 30 out of 40 lowest-energy states for every $l$ up to $l_{max}=5$
are used to calculate $\hat \Sigma$ and 14 lowest states above the
core are used for every $l$ up to $l_{max}=4$ to construct the
two-electron states.

The two-electron valence states are found by solving the eigenvalue
problem,
\begin{equation}
  \hat H^{\rm eff} \Psi_v = E_v \Psi_v \, ,
\label{CI}
\end{equation}
using the standard CI techniques. Calculated and experimental energies of
a few lowest-energy states of Yb are presented in Table~\ref{energies}. One can
see that the pure {\em ab initio} energies are already close to the
experimental values. However, for improving the accuracy further, we
re-scale the correlation operator $\hat \Sigma_1$ by replacing $\hat
\Sigma_1$ in the effective Hamiltonian (\ref{Heff}) in each partial
wave $s, p_{1/2}, p_{3/2} , \ldots$  by $f_a\hat \Sigma_{1}$. The
rescaling factors are $f_s=0.8772$, $f_p=1.03$, $f_d=0.933$, and
$f_f=1$. These values are chosen to fit the experimental spectrum of
Yb. The result of the fitting is shown in Table~\ref{energies}. Note
that the fitting is not exact for all the levels because the number of
levels is larger than the number of fitting parameters.

\begin{table}[h]
\caption{Energy levels of Yb (cm$^{-1}$). The $4f^{13}5d6s^2 (7/2,5/2)^o_1$ level lies outside of the computational model space. 
}
\label{energies}
\begin{tabular}{lccccc}
\hline
\multicolumn{2}{c}{State} & $J$ & Exp.\cite{RareEearhEnergiesBook78} &
\multicolumn{2}{c}{CI+MBPT} \\
              &       &     &      & {\em ab initio} & Rescaled $\hat
              \Sigma_1$ \\
\hline
$4f^{14}6s^2$ & $^1$S     & 0   &     0.0 & 0.0 & 0.0 \\

$4f^{14}6s6p$ & $^3$P$^o$ & 0 & 17288 &  18246 &   17289 \\
             &           & 1 & 17992 &  18946 &   17996 \\
             &           & 2 & 19710 &  20688 &   19759 \\

$4f^{14}5d6s$ & $^3$D     & 1 & 24489 &  24922 &   24489 \\
             &           & 2 & 24752 &  25195 &   24743 \\
             &           & 3 & 25271 &  25765 &   25276 \\[1ex]

\hline
$4f^{14}6s6p$ & $^1$P$^o$ & 1 & 25068 & 26463 &   25611 \\
$4f^{13}5d6s^2$ & $(7/2,5/2)^o$ & 1 & 28857 &   & \\
\hline

$4f^{14}5d6s$ & $^1$D    & 2 & 27678 &  28485 &   27812 \\

$4f^{14}6s7s$ & $^3$S    & 1 & 32695 &  33262 &  32668 \\

$4f^{14}6s7s$ & $^1$S    & 0 & 34351 &  34871 &   34280 \\
\hline
\end{tabular}
\end{table}

Transition amplitudes are found with the random-phase approximation (RPA)
\cite{DzuGin06,DzuFla07}
\begin{equation}
  E1_{vw} = \langle \Psi_v || d_z + \delta V^{N-2} || \Psi_w \rangle, \label{E1}
\end{equation}
where $\delta V^{N-2}$ is the correction to the core potential due to core
polarization by an external electric field. We compile our representative  dipole
matrix elements in Table~\ref{tb:dipoles}. The values were  computed in the CI+MBPT approach.  We also compare with a high-precision result~\cite{TakKomHon04} derived from fitting molecular long-range potentials to photoassociation spectra taken with ultracold samples.
The 16\% theory-experiment disagreement
for the $6s^2\,^1\!S_0   - 6s6p\,^1\!P^o_1$  amplitude is due to mixing of the
 $6s6p \ ^1P^o_1$ state with the core-excited state  $4f^{13}5d6s^2 (7/2,5/2)^o_1$, which lies outside the computational model space (see Section \ref{mixing2}.)

\begin{table}[h]
\caption{Reduced matrix elements (a.u.) of the electric dipole operator for
transitions in Yb. Values were  computed in the CI+MBPT approach. The theory-experiment disagreement
for the $6s^2\,^1\!S_0   - 6s6p\,^1\!P^o_1$ is due to mixing of the
 $6s6p \ ^1P^o_1$ state with the core-excited state  $4f^{13}5d6s^2 (7/2,5/2)^o_1$, which lies outside the computational model space (see Section \ref{mixing2}.) }
\label{tb:dipoles}   
\begin{ruledtabular}
\begin{tabular}{l l l}
\multicolumn{1}{c}{Transition}  &
\multicolumn{1}{c}{Calc.} &\multicolumn{1}{c}{Exp.} \\
\hline

$6s^2\,^1\!S_0   - 6s6p\,^3\!P^o_1$ & 0.587 &       \\
$6s^2\,^1\!S_0   - 6s6p\,^1\!P^o_1$ & 4.825 & 4.148(2)\footnotemark[1] \\
$6s6p\,^3\!P^o_0 - 5d6s\,^3\!D_1$ & 2.911 &       \\
$6s6p\,^3\!P^o_0 - 6s7s\,^1\!S_0$ & 1.952 &       \\
\end{tabular}
\end{ruledtabular}
\noindent \footnotetext[1]{Ref. \cite{TakKomHon04}}
\end{table}

\subsection{Dalgarno-Lewis and RPA methods}
Computing polarizability requires summing over a complete set of {\em many-body} states.
The CI+MBPT subspace of two valence electrons covers only a part of this set.
The other class of intermediate state are core excitations, where the valence electrons
serve as spectators. Notice, that even in the independent particle picture, the state of the valence electrons affects possible
core excitations: core electrons can not be promoted into excited orbitals occupied
by the valence electrons due to the Pauli exclusion principle. Following this discussion,
we divide the polarizability into three parts~\cite{DerJohSaf99},
\[
  \alpha(\omega) = \alpha_\mathrm{val}( \omega) +\alpha_\mathrm{core}( \omega) + \alpha_\mathrm{core-val}( \omega)  \, .
\]
%
%

The valence contribution, $\alpha_\mathrm{val}( \omega)$, is given by (\ref{alphaw}) where $v$ and $k$ are two-electron CI states (i.e., $k$ lie entirely in the model space). We use the Dalgarno-Lewis
method~\cite{DalLew55} for summing over a complete set of two-electron
states. In this method, a correction $\delta \Psi_v$ to the two-electron wave
function of the state $v$ is introduced and the contribution of the valence
electrons to the polarizability is expressed as (here we consider static polarizability)
\begin{equation}
  \alpha_\mathrm{val}(0) = \frac{2}{3}\langle \delta \Psi_v ||\mathbf{d}|| \Psi_v \rangle \, .
\label{eq:deltapsi}
\end{equation}
The correction $\delta \Psi_v$ is found by solving the system of linear inhomogeneous
equations
\begin{equation}
  (\hat H^{\rm eff} - E_v )\delta \Psi_v = - (d_z+\delta V^{N-2}) \Psi_v.
\label{eq:DL}
\end{equation}
In case of dynamic polarizabilities we employ the identity
\[
\frac{E_k-E_v}{(E_k-E_v)^2-\omega^2} =
\frac{1}{(E_k-E_v) - \omega} +
\frac{1}{(E_k-E_v) + \omega}  \, ,
\]
i.e.,  the equations (\ref{eq:DL}) need to be solved twice at two different energies $E=E_v \pm \omega$. For polarizabilities of purely imaginary argument, $\omega \rightarrow i \omega$ in the above equations.

The core polarizability is found using the RPA method\cite{Joh88} (linearized response theory). This method yields required eigen-energies  of many-body particle-hole excitations from the core and relevant electric-dipole transition amplitudes. The dynamic core polarizability reads
\begin{equation}
\alpha_\mathrm{core}( i \omega) =
\sum_{\omega_\mu > 0}
\frac{ f_\mu }
{ \left( \omega_\mu \right)^2 + \omega^2 } \, .
\label{Eqn_acRRPA}
\end{equation}
Here the summation is over particle-hole excitations
from the ground state of the atomic core; $\omega_\mu$ are excitation energies
and $f_\mu$ are the corresponding electric-dipole oscillator strengths.
Technically, compared to the original, differential-equation method of Ref.~\cite{Joh88}, we use a B-spline basis set expansion. An important property of the RPA core polarizability is that it satisfies (non-relativistically) the Thomas-Reiche-Kuhn  sum rule,
$\lim_{\omega \rightarrow \infty} \omega^2 \alpha_\mathrm{core}( i \omega) = N-2$. This property is especially important in evaluating the $C_3$ and $C_6$ coefficients for heavy atoms~\cite{DerJohSaf99}.

Finally, the core-valence counter term, $\alpha_\mathrm{core-val}( \omega)$,  is small. At the DHF level, we find that the relevant contributions to static polarizabilities are below 1\% and we neglect it at our present level of accuracy.


\section{The role of the 
$4f^{14}6s6p \ ^1P^o_1 -  4f^{13}5d6s^2 (7/2,5/2)^o_1$ mixing}

\label{mixing2}

There is a number of relatively low-lying states in the discrete spectrum of Yb which
have an excitation from the outer $4f$ sub-shell of the core.
These states certainly cannot be considered
as states with two valence electrons above closed shells.
One of such states is $4f^{13}5d6s^2 (7/2,5/2)^o_1$ which lies close to the $4f^{14}6s6p \ ^1P^o_1$ state (see Table~\ref{energies}).
Notice that both states have the same total angular momentum and parity.
Therefore the Coulomb interaction strongly mixes the two states.
This mixing  complicates the calculations. Our effective Hamiltonian (\ref{Heff}) does take into account excitations from the core via second-order correlation operator $\hat \Sigma$. Yet, the
second-order approximation is not adequate in this particular case, as
the two states are separated by mere 0.0173 a.u. (3789~cm$^{-1}$).


The calculation of
the ground state polarizability is affected by the $4f^{14}6s6p \
^1P^o_1 -  4f^{13}5d6s^26 (7/2,5/2)^o_1$ mixing.
The $6s^2 \ ^1$S$_0 \rightarrow 6s6p\ ^1$P$^o_1$ electric
dipole transition contributes about 90\% to the polarizability of the
ground state. Corresponding theoretical CI+MBPT transition amplitude,
4.825~a.u., deviates significantly from the experimental value
4.148~a.u.~\cite{TakKomHon04}. As it was pointed out in Ref.~\cite{PorRakKoz99P}
the reason for this disagreement is the above mentioned mixing.

Calculation of the core polarizability displays a similar dependence on the mixing.
Our RPA result for $\alpha_\mathrm{core}(0)=6.39$ a.u. At the same time,
semi-empirical result\cite{ZhaDal07}  based on the life-time measurements of low-lying core-excited
states in Yb  suggests a much higher value of 24(4) a.u.  Again, the reason for
this large difference lies in the strong mixing of the $4f^{13}5d6s^2 (7/2,5/2)^o_1$ and  $4f^{14}6s6p \ ^1P^o_1$ states. Indeed, in the RPA core polarizability calculations, we consider a different system, Yb$^{++}$, which is unaware of the presence of the two valence electrons.

Since the $6s^2 \ ^1$S$_0 \rightarrow 6s6p\ ^1$P$^o_1$ transition
contributes about 90\% to the polarizability of the
ground state, the authors of Ref.~\cite{PorDer06Z} suggested to replace the CI+MBPT
dipole matrix element by its experimental value. This was
done to  improve the accuracy of computing the polarizability. Below we argue that such a substitution
cannot be justified unless a similar
correction is done for the $4f_{7/2} \rightarrow 5d_{5/2}$
transition amplitude in the calculation of the {\em core} polarizability.

Let us use  the following notation for the pure (unmixed)
states
\begin{eqnarray}
  \Phi_1 &=& |4f^{14} 6s6p \ ^1\rm{P}^o_1 \rangle \nonumber \, , \\
  \Phi_2 &=& |4f^{13} 5d6s^2 \ (7/2,5/2)^o_1 \rangle \, . \nonumber
\end{eqnarray}
The mixed states can be written as
\begin{eqnarray}
  \Psi_1 &=& \Phi_1 \cos\phi - \Phi_2 \sin\phi\nonumber \, ,\\
  \Psi_2 &=& \Phi_1 \sin\phi + \Phi_2 \cos\phi,  \nonumber
\end{eqnarray}
where $\phi$ is mixing angle.

Let also use $\Psi_0 \equiv \Phi_0$ for the ground state wave
function.

If we neglect a small energy difference between the states $\Psi_1$ and
$\Psi_2$ (the relevant energy denominators entering $\alpha(0)$ differ by about 10\%) then their total contribution to the polarizability of the
ground state is proportional to
\begin{eqnarray}
\langle \Psi_0||\mathbf{d}|| \Psi_1 \rangle^2 &+&
\langle \Psi_0||\mathbf{d}|| \Psi_2 \rangle^2 = \nonumber \\
\langle \Phi_0||\mathbf{d}|| \Phi_1 \rangle^2 &+&
\langle \Phi_0||\mathbf{d}|| \Phi_2 \rangle^2
\end{eqnarray}
and does not depend on the mixing angle $\phi$. This means that either
bases $\{\Phi_0, \, \Phi_1, \, \Phi_2\}$ or $\{ \Psi_0, \, \Psi_1, \, \Psi_2 \}$ can
be used in the calculations.
The {\em ab initio} calculations within the
CI+MBPT framework in which Yb is considered as a two-valence-electrons atom
uses the non-mixed basis while the analysis based on experimental data
naturally corresponds to the use of the mixed basis. Both approaches
produce close results as will be demonstrated in
Section~\ref{Results}. By contrast, the substitution
of the experimental value for the $\langle 6s^2 \ ^1$S$_0 |\mathbf{D}| 6s6p\
^1$P$^o_1\rangle$ transition amplitude
is equivalent to using the
mixed non-orthogonal basis $\{ \Psi_0, \ \Psi_1, \ \Phi_2 \}$.  This seems to introduce
a large error
into the calculation of the polarizability. Indeed, the value reported in
Ref.~\cite{PorDer06Z,PorDer06BBR} is 111.3 a.u. which is about 30\% smaller
than the results
of most {\em ab initio} calculations and semi-empirical analyses, compiled in Table~\ref{tb:alpha0}. For example,
the analysis~\cite{ZhaDal07} of the experimental data yielded
$\alpha_{^1{\rm S}_0}(0) = 136.4(4.0) \, {\rm a.u.}$

The result of Ref.~\cite{PorDer06Z,PorDer06BBR} is brought into an agreement
with that of Ref.~\cite{ZhaDal07}
by consistently correcting the core polarizability for the mixing.
A  DHF contribution from the $4f$ subshell to core polarizability is 1.9 a.u. If we replace this value by the value~\cite{ZhaDal07} based on experimental data (i.e., account for the mixing),
24(4) then the result of Ref.~\cite{PorDer06Z} increases to
$\alpha_{^1{\rm S}_0}(0)  = 133 \, {\rm a.u.}$ consistent with the value of $136.4(4.0) \, {\rm a.u.}$ from Ref.~\cite{ZhaDal07}.

It is useful to find out what happens if the difference in energies of
the two mixed states is not neglected. We will use a short-hand notation $\Delta E_1=E(\Psi_1)
- E(\Psi_0)$, $\Delta E_2=E(\Psi_2) - E(\Psi_0)$, and $\delta =
1/\Delta E_2- 1/\Delta E_1$. Since the state $\Psi_2 \equiv 4f^{13}5d6s^26
(7/2,5/2)^o_1$ is above the state $\Psi_1 \equiv 4f^{14}6s6p \
^1P^o_1$ the parameter $\delta$ is negative. Further, $A_{01} = \langle \Phi_0||\mathbf{d}|| \Phi_1 \rangle$ and
$A_{02} = \langle \Phi_0||\mathbf{d}|| \Phi_2 \rangle$.

A partial contribution of the two states into the polarizability reads
\begin{equation}
\delta \alpha = \frac{1}{\Delta E_1}(A_{01}^2+A_{02}^2) + \delta
(A_{01}\sin\phi + A_{02}\cos\phi)^2.
\label{eq:delta}
\end{equation}
We see that when the core-excited states lie above the
states with which they are mixed, the correction to the polarizability
due to this mixing is negative. This is the case for both
polarizabilities of Yb considered in this work. We would keep this in
mind while analyzing the accuracy of the calculations in Section~\ref{Results}.

\section{Mixing of  degenerate states}

\label{mixingn}

In the previous section we considered a particular case of mixing of the
$4f^{14}6s6p \ ^1P^o_1$ and  $4f^{13}5d6s^26 (7/2,5/2)^o_1$  states of Yb and
demonstrated that this mixing can be ignored in the {\em ab initio}
calculations of the ground-state polarizability of Yb. Notice that similar
problems arise in computing the polarizability of the $^3$P$^o_0$
state and in many other cases not limited to calculations of the
polarizabilities of Yb. Therefore, it may be useful to treat a general
case and formulate the following theorem:
{\em   Consider a second-order transition amplitude, involving summation
over a complete set of intermediate states. Then,
a contribution from a subspace spanned by degenerate
states does not depend on mixing of these states}.


 A generic contribution to a second-order transition amplitude can be written as (here $E_0$ is some energy off-set)
\begin{equation}
  A_{if} = \sum_m \frac{\langle F_f | \hat P | F_m \rangle
                        \langle F_m | \hat Q | F_i \rangle}{E_0-E_m}
\label{eq:A2}
\end{equation}
Here $F_i$ and $F_f$ are some initial and final atomic states, and $\hat P$ and
$\hat Q$ are some operators. Let's assume that the summation over the complete
set of intermediate states $F_m$ subsumes a summation over a group of
degenerate states $\mathcal{D}$ and rewrite the corresponding partial contribution as
\begin{equation}
  \Delta A_{if} = \frac{1}{\Delta E} \sum_{m \in \mathcal{D}}  \langle F_f | \hat P | F_m \rangle
                        \langle F_m | \hat Q | F_i \rangle.
\label{eq:A2p}
\end{equation}
Here the summation ranges only over degenerate states and $\Delta E$ is their common
energy denominator.

As in the previous section, we introduce the pure non-mixed basis states
$\Phi_k$ and rewrite the real (mixed) states $F_m$ as an expansion
\begin{equation}
  F_m = \sum_k c_{mk} \Phi_k.
\label{eq:Fm}
\end{equation}
Both sets $\Phi$ and $F$ can be made orthonormal, with the matrix $C=\{c_{mk}\}$ being a unitary matrix, i.e., $C C^\dagger=I$.
Substitution of (\ref{eq:Fm}) into (\ref{eq:A2p}) leads to
\begin{eqnarray}
  \Delta A_{if} &=&  \frac{1}{\Delta E} \sum_{kl} \langle F_f | \hat P
  | \Phi_k \rangle \langle \Phi_l | \hat Q | F_i \rangle\sum_m
                        c_{mk}c^*_{ml} \nonumber \\
&=& \frac{1}{\Delta E} \sum_{k} \langle F_f | \hat P | \Phi_k \rangle
                        \langle \Phi_k | \hat Q | F_i \rangle.
\label{eq:theor}
\end{eqnarray}
Here we used the ortho-normality condition, $\sum_k c_{km}c^*_{kl} = \delta_{ml}$.
We see that the final expression does not depend on the mixing coefficients $c$.

We stress two conditions under which this theorem is valid. It is assumed that the unmixed states possess the same energy and that there is no mixing with the states outside the degenerate subspace. These two conditions are probably never fulfilled
exactly. However, there is a large number of cases when they are fulfilled approximately. One of the criteria is that the average energy spread in the quasi-degenerate subspace $|\delta E_\mathcal{D}| \ll |E_0 - \bar{E}_\mathcal{D}|$, where  $\bar{E}_\mathcal{D}$ is the average energy in the subspace.

Why is this practically important? In Yb, to rigorously account for the mixing, one has to extend the two-valence-electron model space to include computationally-expensive core-excited states.
The theorem claims that while computing polarizabilities,
a much smaller two-valence-electron space would suffice.
In other words, at the CI stage,
the excitations from the $4f$ subshell can be safely ignored (even) if there are nearby two-electron states of the same symmetry. (Of course,
{\em virtual} core excitations are included in the self-energy MBPT operator $\Sigma$.)
Excitations from the core will have to be included into calculation of the polarizability of the core. This corresponds to  working with the
non-mixed basis $\Phi_k$.

The independence on mixing explains why pure two-valence {\em ab initio} calculations of polarizabilities give good results for Yb despite its complicated structure. This should
be true not only for polarizabilities but also for two-photon transition
amplitudes, parity non-conservation, etc., and not only for Yb but for
some other atoms as well.

\section{Results and discussion}

\label{Results}

\subsection{Static polarizabilities}

\label{Polarizabilities}
Our computed static polarizabilities of the clock states are presented in Table~\ref{tb:alpha0}. There we tabulate
values obtained in various approximations and we also compile results from the literature.
Below we analyze our results and estimate theoretical uncertainties.

The first line of Table~\ref{tb:alpha0} lists the results of pure {\em
ab initio} calculations, next line gives the results obtained when
energies are fitted as explained in Section~\ref{method}. Then, to gauge the accuracy of the calculations, we
introduce various corrections to the polarizability of the ground
state based on available experimental data. In the third line of the
Table we correct $\alpha_0$ to reproduce the experimental value
for the magic frequency~\cite{BarHoyOat06,BarStaLem08}.
This correction will be explained in the following section.
The fourth value (135.2) is obtained by
replacing the dominant {\em ab initio} contribution by semi-empirical values. This corresponds to
changing from the unmixed $\Phi_1$, $\Phi_2$ basis to the mixed
$\Psi_1$, $\Psi_2$ basis (see Section~\ref{mixing2} for details). Specifically, we
use the value 4.148 a.u.~\cite{TakKomHon04} for the
$\langle 6s^2 \ ^1{\rm S}_0||d|| 6s6p \ ^1{\rm P}_1 \rangle$
transition amplitude and $\delta \alpha = 24(4)$~\cite{ZhaDal07} for
the contribution of the excitations from the $4f$ core subshell. The
resulting value is smaller than both {\em ab initio} results (with and
without fitting) consistent with our observation made at the end of
Section~\ref{mixing2}.
Finally, we rescale  $\alpha(0)$ of the ground state to fit the experimental
value for the van der Waals coefficient $C_6$~\cite{KitEnoKas08}.
Since $C_6$ is simply obtained by integrating $\alpha(i\omega)^2$, Eq.~(\ref{eq:C6}),
and both the integral and the static polarizability are dominated by
the principal transition, the uncertainties of $\alpha(0)$ and
of $C_6$ are correlated (see a detailed discussion in Ref.~\cite{DerPor02Cs}).
The final value for the ground state and its uncertainty, $141(6)$, are chosen
to cover the spread of our numbers in the Table.

Now we turn to the polarizability of the $^3$P$^o_0$ state.
The central point of our final value   is chosen to coincide
with the fitted value. Choosing the lower calculated value as the central point is
justified by the fact that there is a correction due to excitations
for $4f$ and this correction is negative (see the end of
Section~\ref{mixing2} for details). The value of this correction must
be smaller than that for the ground state because corresponding states with
excitations from $4f$ are higher in the spectrum and closer to the
neighboring two-electron states of the same symmetry.
Notice that  the polarizabilities of both clock states end up having a comparable fractional accuracy.
Qualitatively,  this may be explained that while the convergence with respect to increasing basis is slower for the $^3$P$^o_0$ state,
calculations for the $^1$S$_0$ state are affected stronger by the core-excited states.

Results of other calculations and semi-empirical analyses are
presented in Table~\ref{tb:alpha0} for comparison. For the ground state note a good agreement
with the value of Ref.~\cite{ZhaDal07}, $\alpha(0)=136.4(4.0)$, derived using only experimental data. The
value $\alpha(0)=143$  recommended by the same authors is also
consistent with our result.

The result for the ground-state polarizability is in a disagreement with
previous calculations~\cite{PorRakKoz99P,PorDer06Z,PorDer06BBR}.
As discussed in Section~\ref{mixing2} the disagreement for
$^1$S$_0$ comes from  the inconsistent use of experimental matrix element for the
principal transition in these works.  As for the
$^3$P$^o_0$ state,  the present result and that of Ref.~\cite{PorDer06BBR}
differ by about 2 standard deviations and in experimental work this level
of agreement between two values would be acceptable.

Finally, let us emphasize that evaluating accuracy of theoretical calculations is a non-trivial exercise and
due to the lack of high-accuracy experimental data for the  $^3\!P^o_0$ state,
the uncertainty in Ref.~\cite{PorDer06BBR} was estimated as a half of
the difference between pure {\em ab initio} result and the result obtained
with the fitting of the energy. This may be an unreliable estimate.
Suppose we carry out a similar analysis based on the present calculations
for the ground state. Fitting energy in $\Sigma_1$ moves our {\em ab initio} result, 138.9,
to 145.7, so using prescription of  Ref.~\cite{PorDer06BBR} would yield
145.7(3.4) which is displaced compared to and is as twice as accurate as our final value of 141(6).


\begin{table*}[h]
\caption{Static electric-dipole polarizability $\alpha(0)$ (in a.u.) of the $6s^2 \, ^1\!S_0$ and $6s6p \, ^3\!P^o_0$ clock states of Yb atom in various approximations and comparison with  literature values.}
\label{tb:alpha0}
\begin{ruledtabular}
\begin{tabular}{l l l}
$\alpha(0), ^1$S$_0$ &$\alpha(0), ^3$P$^o_0$ & Comment \\
\hline
\multicolumn{2}{c}{this work} & \\
138.9 & 315.9 & {\em ab initio} with no fitting \\
145.7 & 301.6 & energies are fitted by rescaling $\hat \Sigma_1$ \\
139.1 &      & with a correction to $\alpha(^1{\rm S}_0)$
            to fit the magic frequency $\omega=0.0600$ a.u. \\
135.2 &     & using experimental data for dominant contributions \\
138.8 &     & rescaled to fit experimental data\cite{KitEnoKas08} for $C_6 = 1932(30)$ a.u.
\\

141(6) & 302(14) & final \\
\hline
\multicolumn{2}{c}{other} & \\
131.6 &  &  Wang {\em et al}~\cite{WanPanSch95}; Density functional theory, 1995 \\
145.3 &  &  Wang and Dolg~\cite{WanDol98}; CCSD(T), 1998 \\
141.7 &  &  Miller~\cite{CRC04}; Density functional theory, 2002\\

118(45) & 252(25) & Porsev {\em et al}~\cite{PorRakKoz99P}; CI+MBPT, 1999 \\
111.3(5) & 266(15) & Porsev and Derevianko~\cite{PorDer06Z,PorDer06BBR};
CI+MBPT, 2006 \\
154.7 &  & Buchachenko {\em et al}~\cite{BucSzcCha06}; CCSD(T), 2006 \\
136.4(4.0) & & Zhang and Dalgarno~\cite{ZhaDal07}; based on
              experimental data, 2007 \\
157.3 &  &  Chu  {\em et al}~\cite{ChuDalGro07}; Density functional theory, 2007 \\
143 & & recommended value by Zhang and Dalgarno~\cite{ZhaDal07}, 2007\\
144.59 &  & Sahoo and Das~\cite{SahDas08}; relativistic coupled-cluster
theory, 2008 \\
\end{tabular}
\end{ruledtabular}
\end{table*}

\subsection{Magic frequencies}


Magic frequencies are frequencies of the laser  field at which
the ac Stark shifts of both clock levels are the same so that
the frequency of the clock transition is insensitive to the laser intensity. Magic frequencies, $\omega^*$, are found from the condition
\begin{equation}
  \alpha_{^1{\rm S}_0}(\omega^*) = \alpha_{^3{\rm P}_0}(\omega^*).
\label{eq:magic}
\end{equation}
First five calculated magic frequencies for Yb are presented in
Table~\ref{tb:magic}. The first frequency was measured to a high
precision~\cite{BarHoyOat06,BarStaLem08}. Our {\em ab initio} calculations
reproduce it with the 1.3\% accuracy. The accuracy for the other 
frequencies is likely to be worse. This is because the energy denominators
for the dynamic polarizabilities (\ref{alphaw}) are shifted by the
frequency of the laser field and the assumption that the difference
in energies of the $4f^{14}6s6p \ ^1P^o_1$ and$ 4f^{13}5d6s^26
(7/2,5/2)^o_1$ states can be neglected becomes less accurate (see
Section~\ref{mixing2} for details).

To improve the accuracy of predicting the magic
frequencies and corresponding polarizabilities
we modify formula (\ref{alphaw}) in two
ways. First, we use experimental data for the dominant terms:
$\langle 6s^2 \ ^1{\rm S}_0 ||d||6s6p \ ^1{\rm
  P}_1\rangle$=4.148~\cite{TakKomHon04} and $\langle 6s^2 \ ^1{\rm S}_0
||d||4f^{13}5d6s^2 \  (7/2,3/2)^o_1 \rangle$=2~\cite{BlaKom94}.
Second, we introduce a correction to simulate the effect of other
excitations from the $4f$ core shell,
\begin{equation}
  \Delta \alpha =
  \frac{A^2}{3}\left(\frac{1}{E-\omega}+\frac{1}{E+\omega}\right),
\label{eq:deltaal}
\end{equation}
where $E=37414.59$ cm$^{-1}$ is the energy of the nearby state with the
excitation from the core and $A$ is a fitting parameter. Its value is
chosen to fit experimental magic frequency. The rest of contributions
are taken from the {\em ab initio} calculations.

The above empirical correction has been introduced for the ground state only.
The polarizability of the upper clock state was computed within the CI+MBPT method
 for the valence contribution and the DHF method for core polarizability.

\begin{figure}[h]
\begin{center}
\includegraphics*[scale=0.5]{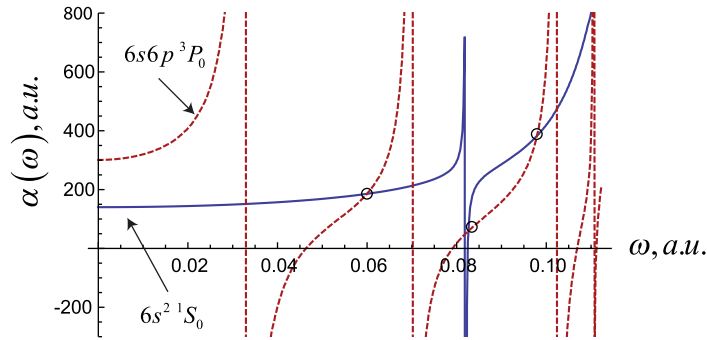}
\caption{(Color online) Dynamic polarizabilities of the two clock levels. ``Magic'' conditions occur when the two polarizabilities intersect off resonance. These are marked by small circles on the plot.
\label{Fig:Magic}}
\end{center}
\end{figure}

The values of magic
frequencies and corresponding polarizabilities calculated with this
empirical correction are listed in the third and forth columns of
Table~\ref{tb:magic} and the ac polarizabilities are shown in Fig.\ref{Fig:Magic}. 
 We see that the results for the first four magic
frequency change little, while those for the fifth magic frequency vary
significantly.
Still, while difficult to predict, the fifth magic frequency
 may be of a practical interest for designing better clocks: by contrast to the first four frequencies, the ac polarizability here may be  negative (that was the {\em ab initio} prediction, but the empirical correction flipped the sign of the polarizability). Because of that, the atoms are trapped
at the minima of the laser intensity, which reduces perturbations of the
clock frequency~\cite{TakKatMar09}.

Finally, using the modified formula at $\omega=0$ provides another accuracy test for our predicted
static polarizability of the ground state (Section~\ref{Polarizabilities}). The resulting value, 139.1
a.u. was listed in Table~\ref{tb:alpha0} and used in
estimating the accuracy of the calculations.

\begin{table}
\caption{Magic wavelengths for the  $6s^2 \, ^1\!S_0 \rightarrow  6s6p \, ^3\!P^o_0$
  transition in Yb and  values of dynamic polarizabilities at corresponding frequencies (a.u.).}
\label{tb:magic}
\begin{ruledtabular}
\begin{tabular}{l l l l l}
\multicolumn{2}{c}{\em ab initio} &
\multicolumn{2}{c}{Corrected\footnotemark[1]} & Experiment\\
\multicolumn{1}{c}{$\lambda$,nm} & \multicolumn{1}{c}{$\alpha(\omega)$,a.u.} &
\multicolumn{1}{c}{$\lambda$,nm} & \multicolumn{1}{c}{$\alpha(\omega)$,a.u.} &
\multicolumn{1}{c}{$\lambda$,nm} \\
\hline
749.0 & 193 & 759.37 & 186 & 759.355 \footnotemark[2] \\
552.2 & 46 & 551.5 & 60 & \\
459.3 & 478 & 465.4 & 382 & \\
413.2 & 1246 &  413.25 & 817 \\
359.7 & -742 &  402.55\footnotemark[3] & 365\footnotemark[3]  \\
\end{tabular}
\end{ruledtabular}
\noindent \footnotetext[1]{$\alpha_{^1{\rm S}_0}(\omega)$ is
corrected to fit experimental magic frequency}
\noindent \footnotetext[2]{Ref. \cite{BarStaLem08}}
\noindent \footnotetext[3]{unreliable}
\end{table}

\subsection{Black-body radiation shift}

The effect of black-body radiation (BBR) on the frequency of the Yb clock
transition  is expressed in terms of the static polarizabilities.
The relevant  clock shift reads (see detailed derivations in Ref.~\cite{PorDer06BBR})
\begin{equation}
\delta \nu_\mathrm{BBR} \approx -\frac{2}{15} (\alpha_\mathrm{hfs} \pi)^3 \,  T^4
\left(\alpha_{^3\!P_0}(0) - \alpha_{^1\!S_0}(0)\right), \label{Eq:BBR}
\end{equation}
where $T$ is the temperature and $\alpha_\mathrm{hfs}$ is the fine-structure constant.
With our polarizabilities we find at $T=300$ K, $\delta \nu_{\rm BBR}=-1.41(17)$~Hz which is
in good agreement with the result, $\delta \nu_{\rm BBR}=-1.34(13)$~Hz,
of Ref.~\cite{PorDer06BBR} despite our significant differences in values of individual
polarizabilities. The agreement is fortuitous, as
 both polarizabilities of Ref.~\cite{PorDer06BBR} are smaller, most of
the shift canceling out while taking the difference in Eq.(\ref{Eq:BBR}). Note, however, that a somewhat smaller uncertainty reported in \cite{PorDer06BBR} is mostly due to overestimating the theoretical accuracy for polarizabilities, especially for the ground state. See Sections
\ref{mixing2} and \ref{Polarizabilities} for details.

It is worth emphasizing, that at the time of writing, the BBR correction is the leading uncertainty in the error budget of the lattice clocks~\cite{LemLudBar09}. The present calculations demonstrate a difficulty of atomic theory in determining this important correction accurately. We hope that our analysis would motivate a further experimental work, for example, on cryogenically cooled clock chambers to reduce (and ultimately to accurately determine) the BBR shift.

\subsection{Atom-wall interaction coefficients $C_3$ and van der Waals
  coefficients $C_6$}

We calculate the atom-wall interaction constants $C_3$ and the van der Waals
constant $C_6$ for both clock states using the formulae (\ref{eq:C3})
and (\ref{eq:C6}). The results are presented in Table~\ref{tb:c3c6}
together with the results of other calculations and measurements for
the $C_6$ coefficient for the ground state.

For two Yb atoms in their ground states we find
$C_6(^1\!S_0+^1\!S_0)=2041\, \mathrm{a.u.}$ (CI+MBPT with rescaled $\Sigma$ to fit the energies). However, from our calculations of the polarizabilities
we know that the contributions from the excitations from the $4f$ core
state are likely to make the results smaller. Therefore, we rescaled the central
point for $C_6(^1\!S_0+^1\!S_0)$ at twice the rate as for the polarizability of
the ground state (i.e., reduced it by 6.5\%). This moves the the central value to 1909 a.u.  
Since the $C_6$ is obtained as a quadrature of the square of  polarizability, we assign a fractional uncertainty to $C_6$
as twice as large as that for the static polarizabilities.
Note a good agreement of the ground-state $C_6$ with recent experimental determinations~\cite{KitEnoKas08} and with
calculations~\cite{ZhaDal07}. The results in the Table for  $C_6(^1\!S_0+^3\!P_0)$ and
$C_6(^3\!P_0+^3\!P_0)$ were obtained in CI+MBPT with rescaled $\Sigma$ to fit the energies.

Including core polarizability into computation of $C_6$ is important. We find
that neglecting $\alpha_\mathrm{core}(i \omega)$ reduces the final result for $C_6$
by about 10\%. This is consistent with the earlier observation~\cite{DerJohSaf99}
for heavy alkali-metal atoms.

Atom-wall interaction coefficients $C_3$  for the clock states
were obtained in CI+MBPT with rescaled $\Sigma$ to fit the energies; these
are also listed in the Table~\ref{tb:c3c6}. Notice, that
the difference in $C_3$ coefficients for the clock states
was computed by us earlier in Ref.\cite{DerObrDzu09}. While forming
the difference the contribution of core polarizability cancels out.
For individual states, this contribution is substantial. For example,
for the ground state result is increased from 2.09 to 3.34 a.u. due
to $\alpha_\mathrm{core}(i \omega)$. As a large fraction of $C_3$ value is accumulated due to the core excitations, assigning error to $C_3$ is difficult
For example, to estimate error in $C_3$ for simpler alkali-metal atoms in Ref.~\cite{DerJohSaf99}, results of two methods of computing $C_3$, (i) via the integral of $\alpha(i \omega)$ and (ii) by computing expectation value of some two-body dipole-dipole interaction were compared. A highly-technical evaluation of the expectation value of two-body operator is beyond the scope of the present paper and we do not assign uncertainties to $C_3$.
We expect that the errors in $C_3$ are unlikely to exceed 50\%.

\begin{table}[h]
\caption{Atom-wall interaction coefficients $C_3$ and van der Waals
  coefficients $C_6$ (a.u.) for the  $6s^2 \, ^1\!S_0$ and $6s6p \, ^3\!P^o_0$ states of Yb.   }
\label{tb:c3c6}
\begin{ruledtabular}
\begin{tabular}{l l l l l}
\multicolumn{3}{c}{Value} &
\multicolumn{2}{c}{Comment} \\
\hline
\multicolumn{5}{c}{Atom-wall interaction} \\
$C_3$ & $^1\!S_0$   & 3.34 & Calc. & this work \\
$C_3$ & $^3\!P^o_0$ & 3.68 & Calc. & this work \\
\hline
\multicolumn{5}{c}{Atom-atom interaction} \\
$C_6$ & $^1\!S_0 + ^1\!S_0$   & 1909(160)  & Calc. & this work \\
      &             & 2300(250) & Exp. & Enomoto {\em et al}~\cite{EnoKitKas07} \\
      &             & 1932(35)  & Exp. & Kitagawa {\em et al}~\cite{KitEnoKas08} \\
      &             & 2400 - 2800 & Calc. & Buchachenko {\em et al}~\cite{BucSzcCha06} \\
      &             & 2292  & Calc. & Chu {\em et al}~\cite{ChuDalGro07} \\
      &             & 2062  & Calc. & Zhang and Dalgarno~\cite{ZhaDal07} \\
$C_6$ & $^3\!P^o_0 + ^3\!P^o_0$ & 3886(360)  & Calc. & this work \\
$C_6$ & $^1\!S_0 + ^3\!P^o_0$ & 2709(338)  & Calc. & this work \\
\end{tabular}
\end{ruledtabular}
\end{table}

\section{Conclusion}

We  calculated polarizabilities of the clock states of Yb with the
accuracy of about 5\%. The substantial disagreement with previous calculations for the
ground state polarizability was explained and resolved. We also computed the first four
magic frequencies of the lattice laser field, the effect of black-body
radiation on the frequency of the clock transition, the $C_3$ atom-wall
interaction constants and the $C_6$ van der Waals coefficients for both
clock states. Polarizabilities, the first magic frequency and the $C_6$
coefficient for the ground state are in a good agreement with the most
accurate calculations and measurements.
The presented data may be of interest for designing better clocks,
 applications of the clocks in studying atom-wall interaction and quantum information processing, and quantifying molecular potentials for ultracold collision studies.

\acknowledgements

The authors are grateful to S. G. Porsev and A. Dalgarno for useful discussions.
The work was supported in part by the Australian Research Council and U.S. National Science Foundation.


\end{document}